\documentclass[letter,twocolumn]{jpsj3}
\usepackage{bm}

\def\calH{\mathcal{H}}

\def\ave#1{\langle#1\rangle}   
\def\gsim{\,$\raise0.3ex\hbox{$>$}\llap{\lower0.8ex\hbox{$\sim$}}$\,}
\def\lsim{\,$\raise0.3ex\hbox{$<$}\llap{\lower0.8ex\hbox{$\sim$}}$\,}

\title{Haldane, Large-$D$ and Intermediate-$D$ States
in an $S\!=\!2$ Quantum Spin Chain
with On-Site and $XXZ$ Anisotropies}

\author{Takashi Tonegawa\thanks{E-mail address: tone0115@vivid.ocn.ne.jp},
Kiyomi Okamoto$^{1}$, Hiroki Nakano$^{2}$, T\^oru Sakai$^{2,3,4}$,
Kiyohide Nomura$^{5}$\\
and Makoto Kaburagi}
\inst{Professor Emeritus, Kobe University, Kobe 657-8501, Japan\\
$^{1}$Department of Physics, Tokyo Institute of Technology, Meguro-ku, Tokyo
152-8551, Japan \\
$^{2}$Graduate School of Material Science, University of Hyogo, Hyogo
678-1297, Japan \\
$^{3}$Japan Atomic Energy Agency, SPring-8, Hyogo 679-5148, Japan\\
$^{4}$Transformative Research-Project on Iron Pnictide, JST, Saitama
332-0012, Japan \\
$^{5}$Department of Physics, Kyushu University, Fukuoka 812-8581, Japan
}
\abst{
Using mainly numerical methods, we investigate the ground-state phase diagram
of the \hbox{$S\!=\!2$} quantum spin chain described by
\hbox{$\calH\!=\!\sum_j (S_j^x S_{j+1}^x\!+\!S_j^y S_{j+1}^y\!+\!\Delta S_j^z S_{j+1}^z)\!+\!D \sum_j (S_j^z)^2$},
where $\Delta$ denotes the $XXZ$ anisotropy parameter of the nearest-neighbor
interactions and $D$ the on-site anisotropy parameter.  We restrict ourselves
to the case of \hbox{$\Delta\!\ge\!0$} and \hbox{$D\!\ge\!0$} for
simplicity.  Each of the phase boundary lines is determined by the level
spectroscopy or the phenomenological renormalization analysis of numerical
results of exact-diagonalization calculations.  The resulting phase diagram on
the $\Delta$-$D$ plane consists of four phases; the $XY$1 phase, the
Haldane/large-$D$ phase, the intermediate-$D$ phase and the N\'eel
phase. The remarkable natures of the phase diagram are as follows:~(1)
    there exists an intermediate-$D$ phase which was predicted by Oshikawa in
    1992;
(2) the Haldane state and the large-$D$ state belong to the same phase;
(3) the shape of the phase diagram on the $\Delta$-$D$ plane is different from
    that considered so far.
We note that this is the first report on the observation of the
intermediate-$D$ phase.
}

\kword{$S\!=\!2$ quantum spin chain, ground-state phase diagram, quantum phase
transition}

\begin{document}
\maketitle


Quantum spin chain systems have been attracting increasing attention in recent
years because they provide rich physics even when models are rather
simple.  The most well-known example may be the existence of the Haldane
phase\cite{haldane83a,haldane83b} in integer spin chain systems only with
isotropic nearest-neighbor (nn) interactions.

In this study, using mainly numerical methods, we explore the ground-state
phase diagram of the \hbox{$S\!=\!2$} quantum spin chain described by the
Hamiltonian
\begin{equation}
    \calH
    = \sum_j (S_j^x S_{j+1}^x + S_j^y S_{j+1}^y + \Delta S_j^z S_{j+1}^z)
            + D \sum_j (S_j^z)^2\,,
    \label{eq:ham}
\end{equation}
where $S_j^\alpha$ (\hbox{$\alpha=x$}, $y$, $z$) is the $\alpha$-component of
the $S\!=\!2$ operator at the $j$-th site, and $\Delta$ and $D$ are,
respectively, the $XXZ$ anisotropy parameter of the nn interactions and the
on-site anisotropy parameter.  Hereafter, we denote the total
number of spins in the system by $N$, assumed to be even, and the $z$
component of the total spin by \hbox{$M\bigl(=\!\sum_j S_j^z\bigr)$}.

\begin{figure}[h]
   \begin{center}
       \scalebox{0.2}{\includegraphics{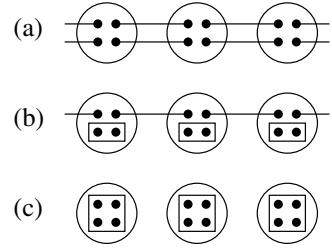}}
   \end{center}
   \caption{Valence bond pictures of (a) the Haldane state, (b) the ID state
   and (c) the LD state.  Large circles denote \hbox{$S\!=\!2$} spins and dots
   \hbox{$S\!=\!1/2$} spins.  Solid lines represent valence bonds
   $\bigl($singlet pairs of two \hbox{$S\!=\!1/2$} spins,
   $(1/\sqrt{2})(\uparrow\downarrow-\downarrow\uparrow)$$\bigr)$.  Two
   \hbox{$S\!=\!1/2$} spins in rectangles are in the
   \hbox{$(S_{\rm tot},S^z_{\rm tot})\!=\!(1,0)$} state and four
   \hbox{$S\!=\!1/2$} spins in squares are in the
   \hbox{$(S_{\rm tot},S^z_{\rm tot})\!=\!(2,0)$} state.}
   \label{fig:vbs-pictures}
\end{figure}
\begin{figure}[h]
   \begin{center}
       \scalebox{0.23}{\includegraphics{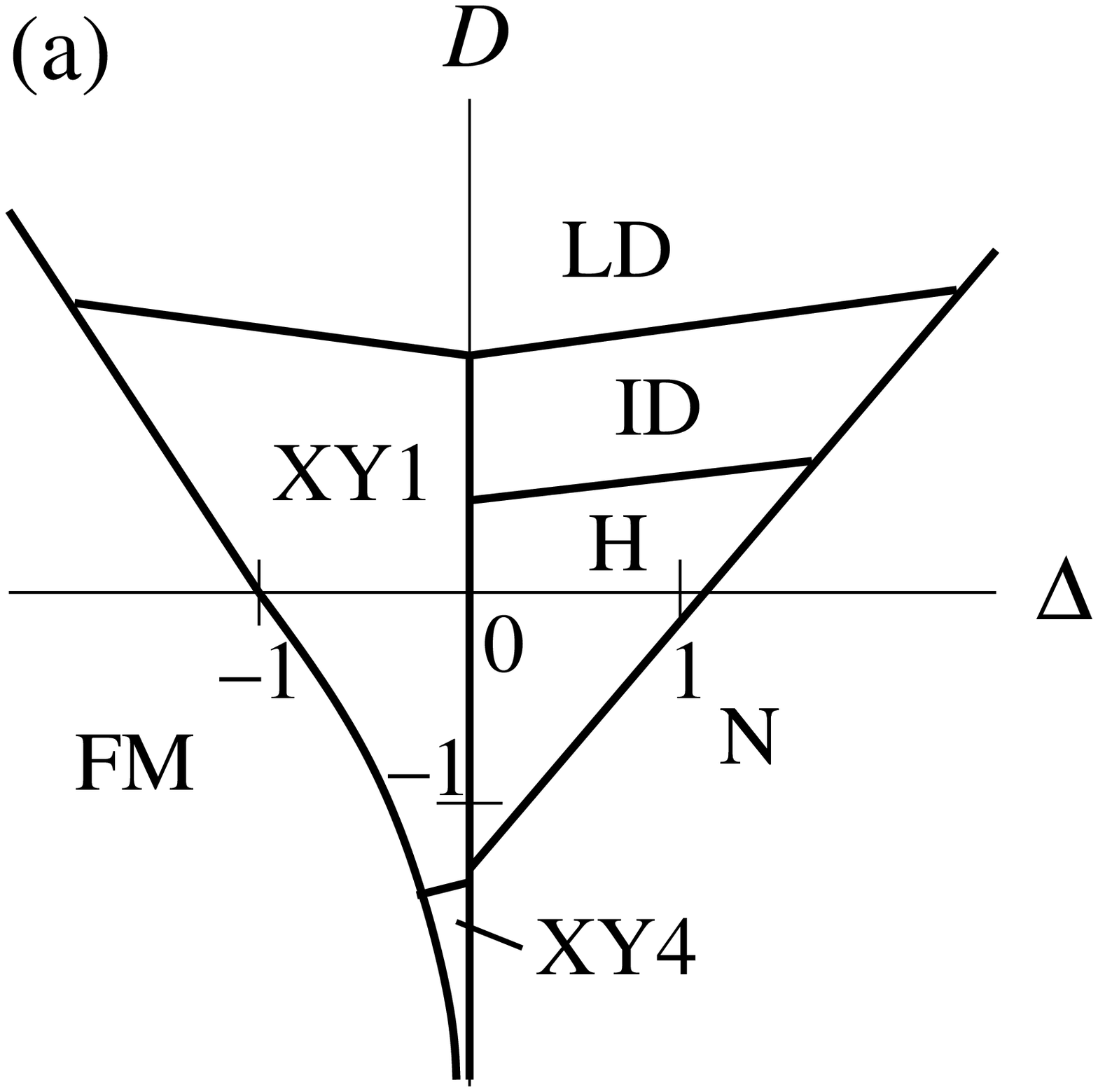}}~~~~
       \scalebox{0.23}{\includegraphics{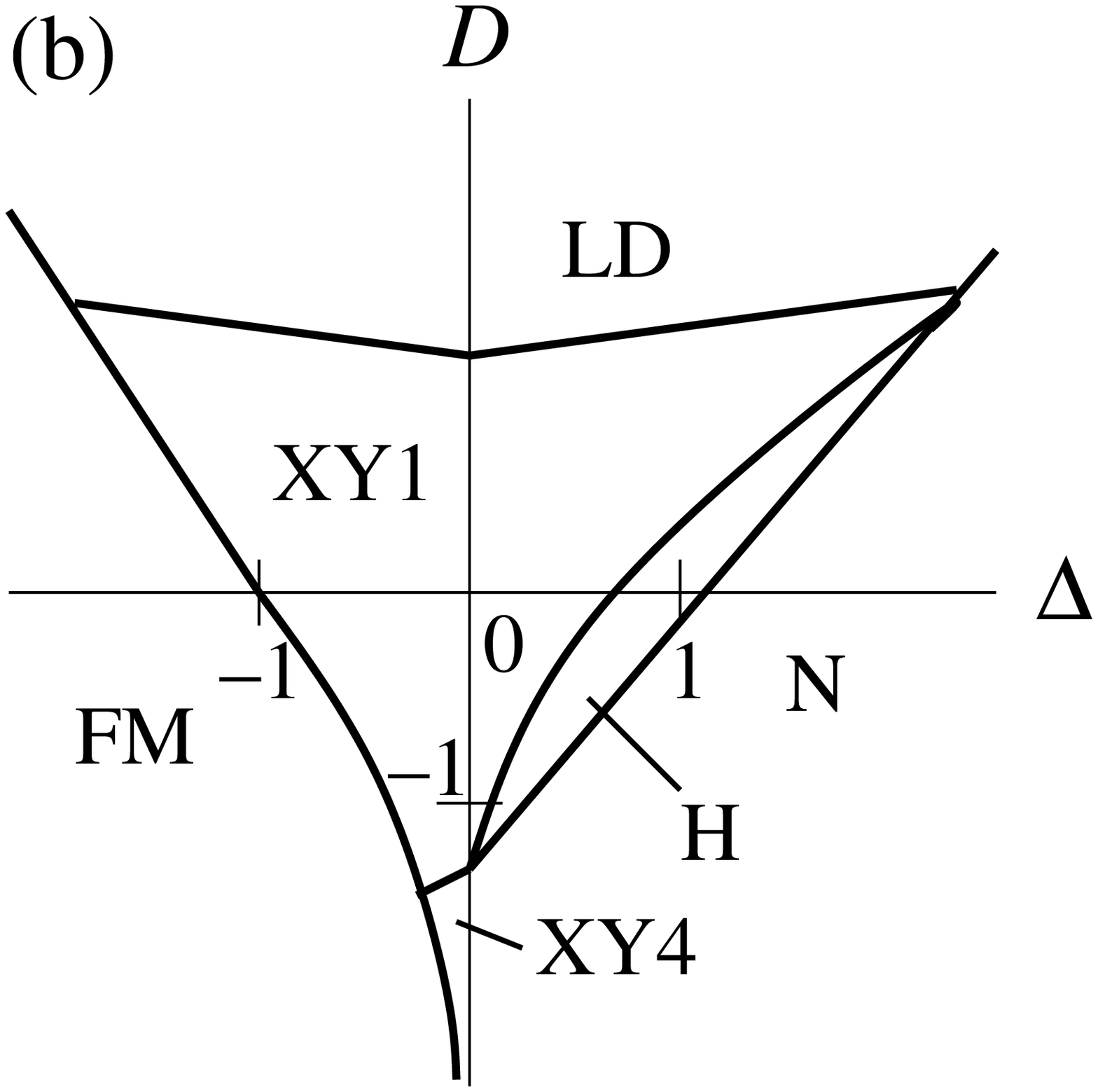}}
   \end{center}
   \caption{(a) The shape of the ground-state phase diagram of the Hamiltonian
            (\ref{eq:ham}), predicted by Oshikawa\cite{oshikawa92} and
            interpreted by Aschauer and Schollw\"ock.\cite{aschauer}  
            The symbols FM, XY1, XY4, LD, ID, H and N stand, respectively,
            for the ferromagnetic, XY1, XY4, large-$D$, intermediate-$D$,
            Haldane and N\'eel phases.  The boundary lines between the $XY1$
            phase and the ID or H phase are not necessary to be
            \hbox{$\Delta\!=\!0$}.  In other words, the $XY$1 phase may intrude
            into the \hbox{$\Delta\!>\!1$} region.
            (b) The shape of the ground-state phase diagram of the Hamiltonian
            (\ref{eq:ham}), proposed by Schollw\"ock et
            al.\cite{aschauer,schollwoeck1,schollwoeck2} by the DMRG
            calculation.
   }
   \label{fig:prediction}
\end{figure}

The ground-state phase diagram on the $\Delta$-$D$ plane of the same
Hamiltonian in the \hbox{$S\!=\!1$} case was discussed by Schulz\cite{schulz}
and den Nijs and Rommels,\cite{den-nijs} and numerically determined by Chen et
al.\cite{chen}.  In the present \hbox{$S\!=\!2$} case, 
Schulz\cite{schulz} discussed the ground-state phase diagram with six phases
obtained by the bosonization method; the phases are, in our terminology, the
ferromagnetic (FM) phase, the N\'eel phase, the $XY$1 phase,
the $XY$4 phase, the Haldane phase and the large-$D$ (LD) phase. The $XY$1
state is characterized by the power decay of the spin correlation function
\hbox{$G_{\perp 1}(r)\!\equiv\!\ave{S_1^+ S_{1+r}^-}$} and the exponential
decay of \hbox{$G_{\perp 4}(r)\!\equiv\!\ave{\{S_1^+)\}^4 \{S_{1+r}^-\}^4}$},
whereas the $XY$4 state by the exponential decay of $G_{\perp 1}(r)$
and the power decay of $G_{\perp 4}(r)$.  The valence bond pictures of the
Haldane state and the LD state are shown in Figs.~\ref{fig:vbs-pictures}(a)
and (c), respectively.  About twenty years ago, Oshikawa\cite{oshikawa92}
predicted, in \hbox{$S\!\ge\!2$} integer quantum spin cases, the existence of
the intermediate-$D$ (ID) phase, the valence bond picture of which is depicted
in Fig.~\ref{fig:vbs-pictures}(b).  Figure \ref{fig:prediction}(a) is an
interpretation of Oshikawa's prediction by Aschauer and
Schollw\"ock\cite{aschauer} in the \hbox{$S\!=\!2$} case.

On the other hand, carrying out the density-matrix renormalization-group (DMRG)
calculation, Schollw\"ock et al.\cite{schollwoeck1,schollwoeck2} and Aschauer
and Schollw\"ock\cite{aschauer} proposed the phase diagram in
Fig.~\ref{fig:prediction}(b) and concluded the absence of the ID phase in the
present \hbox{$S\!=\!2$} model.  Furthermore, performing the level
spectroscopy\cite{kitazawa,nomura} (LS)
analysis of the numerical results of exact-diagonalization calculations, Nomura
and Kitazawa\cite{nomura} showed in the case of $\Delta=1$ that, with the
increase of $D$ from zero, the ground state changes from the Haldane state to
the $XY$1 state at \hbox{$D_{\rm c1}\!=\!0.043$} and after that from the $XY$1
state to the LD state at \hbox{$D_{\rm c2}\!=\!2.39$}.  Thus, it has been
considered for a long time that the ID phase does not exist in the phase
diagram of this model.

\begin{figure}[h]
   \begin{center}
       \scalebox{0.28}{\includegraphics{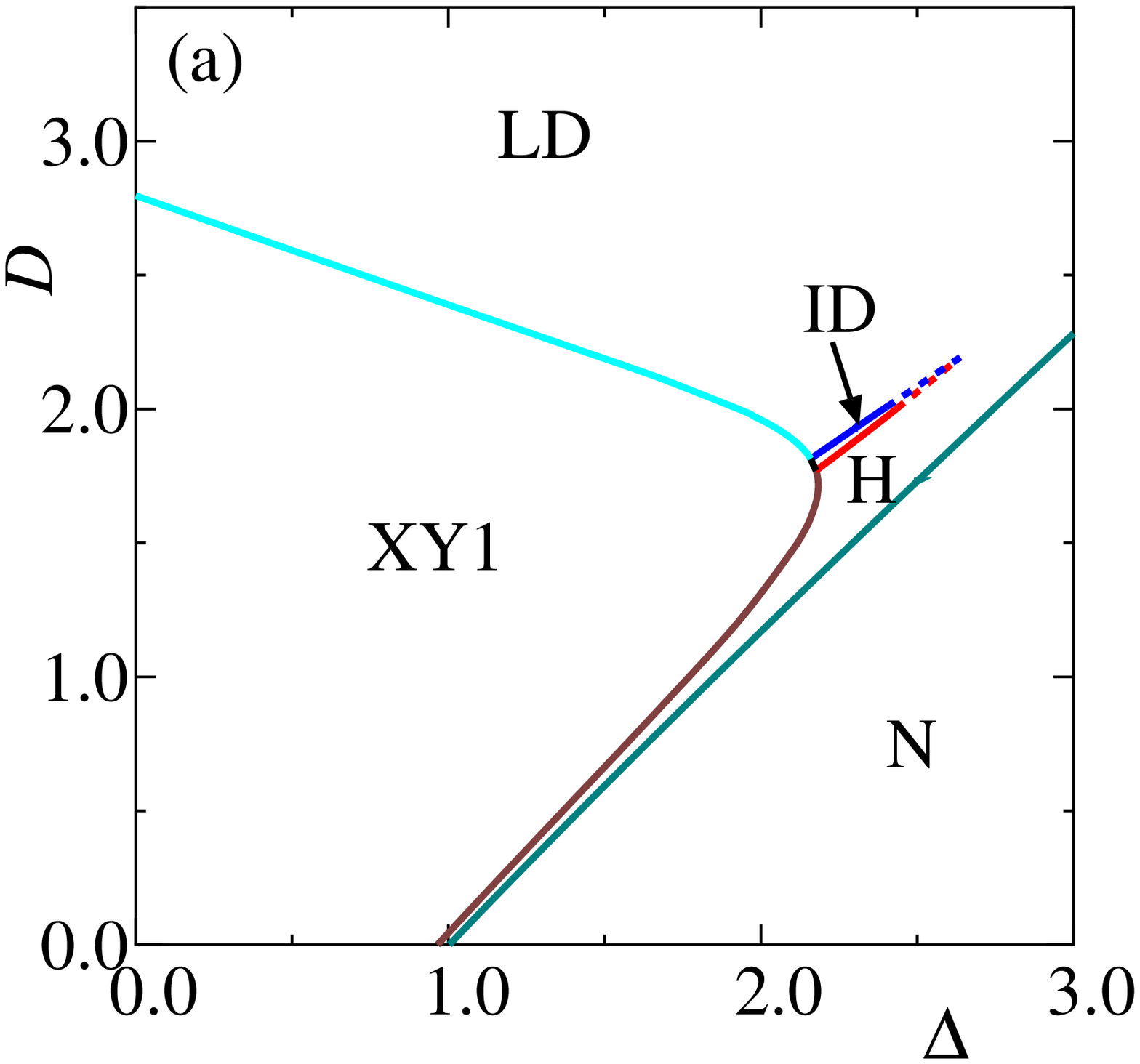}}
   \end{center}
   \begin{center}
       \scalebox{0.28}{\includegraphics{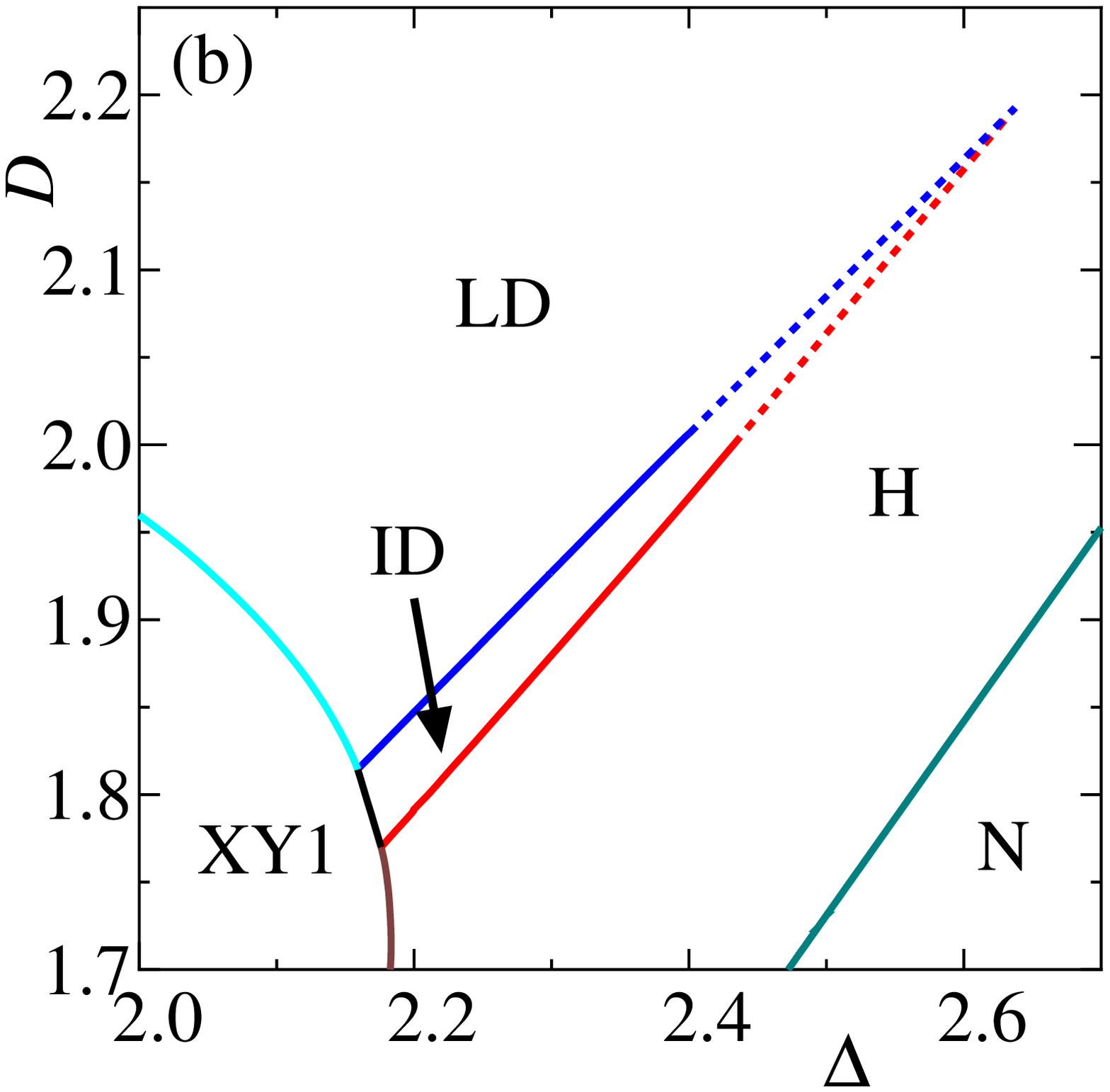}}~~~~
   \end{center}
   \caption{(Color) The ground-state phase diagram on the $\Delta$-$D$ plane
            determined in the present work; in (b), part of (a) is
            enlarged.  Note that the H and LD states belong to the same
            phase.  See the text for more details.
            }
   \label{fig:phasediagram}
\end{figure}
The purpose of the present work is to determine as precisely as possible the
ground-state phase diagram on the $\Delta$-$D$ plane of the \hbox{$S\!=\!2$}
quantum spin chain described by the Hamiltonian (\ref{eq:ham}).  
We mainly employ numerical methods based on the exact-diagonalization
calculation for finite-spin systems with up to \hbox{$N\!=\!12$}, the details
of which are described below.  For simplicity, we restrict ourselves to the
\hbox{$\Delta\!\ge\!0$} and \hbox{$D\!\ge\!0$} case.  The full phase diagram
will be discussed in our future reports.

Figure~\ref{fig:phasediagram} shows our final result for the ground-state
phase diagram.  In this phase diagram five states, that is, the $XY$1, LD, ID,
Haldane and N\'eel states, appear as the ground states.  As we will discuss
later in detail, the Haldane and LD states belong to the same phase, which we
denote as the Haldane/large-$D$ (Haldane/LD) phase.  Therefore, the phase
diagram consists of four phases; they are the $XY$1, LD, Haldane/ID and
N\'eel phases.

In the following discussions for the determination of phase boundary lines,
we denote, respectively, by $E_0(N,M;{\rm pbc})$ and $E_1(N,M;{\rm pbc})$ the
lowest and second-lowest energy eigenvalues of the Hamiltonian (\ref{eq:ham})
with periodic boundary conditions within the subspace determined by $N$ and
$M$.  We also denote by $E_0(N,M,P;{\rm tbc})$ the lowest energy
eigenvalue of the Hamiltonian (\ref{eq:ham}) with twisted boundary conditions,
where an antiferromagnetic bond is twisted, within the subspace determined by
$N$, $M$ and $P$.  Here, $P(=\!+1$ or $-1)$ is the eigenvalue of the space
inversion operator with respect to the twisted bond.  

Both the transition between the Haldane and ID states and that between the ID
and LD states are the Gaussian transition,
and those
between the $XY$1 state to one of the above three states are the
Berezinskii-Kosterlitz-Thouless transition\cite{BKT1,BKT2}.
Following the LS method\cite{kitazawa,nomura},
we should compare three excitation energies,
$E_0(N,0,+1;{\rm tbc}) - E_0(N,0;{\rm pbc})$,
$E_0(N,0,-1;{\rm tbc})- E_0(N,0;{\rm pbc})$ and
$E_0(N,2;{\rm pbc}) - E_0(N,0;{\rm pbc})$ in the
\hbox{$N\!\to\!\infty$} limit.
Namely, the ground state is one of the Haldane/LD, ID and $XY$1 states,
depending on whether the first, second or third excitation energy is the lowest
among them.  A physical and intuitive explanation for this method
is given in our recent paper.\cite{okamoto}

\begin{figure}[h]
   \begin{center}
       \scalebox{0.2}{\includegraphics{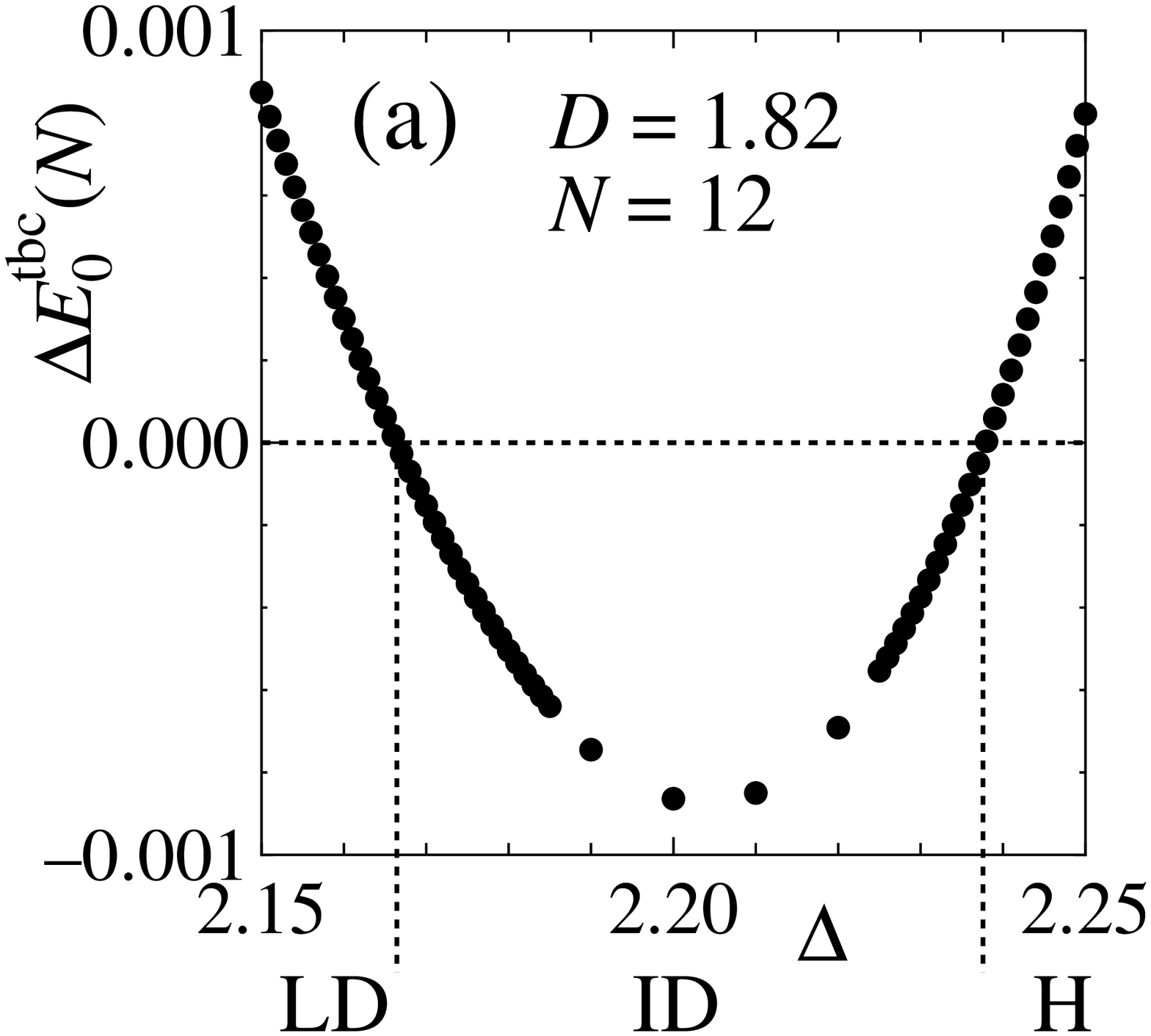}}
       \scalebox{0.2}{\includegraphics{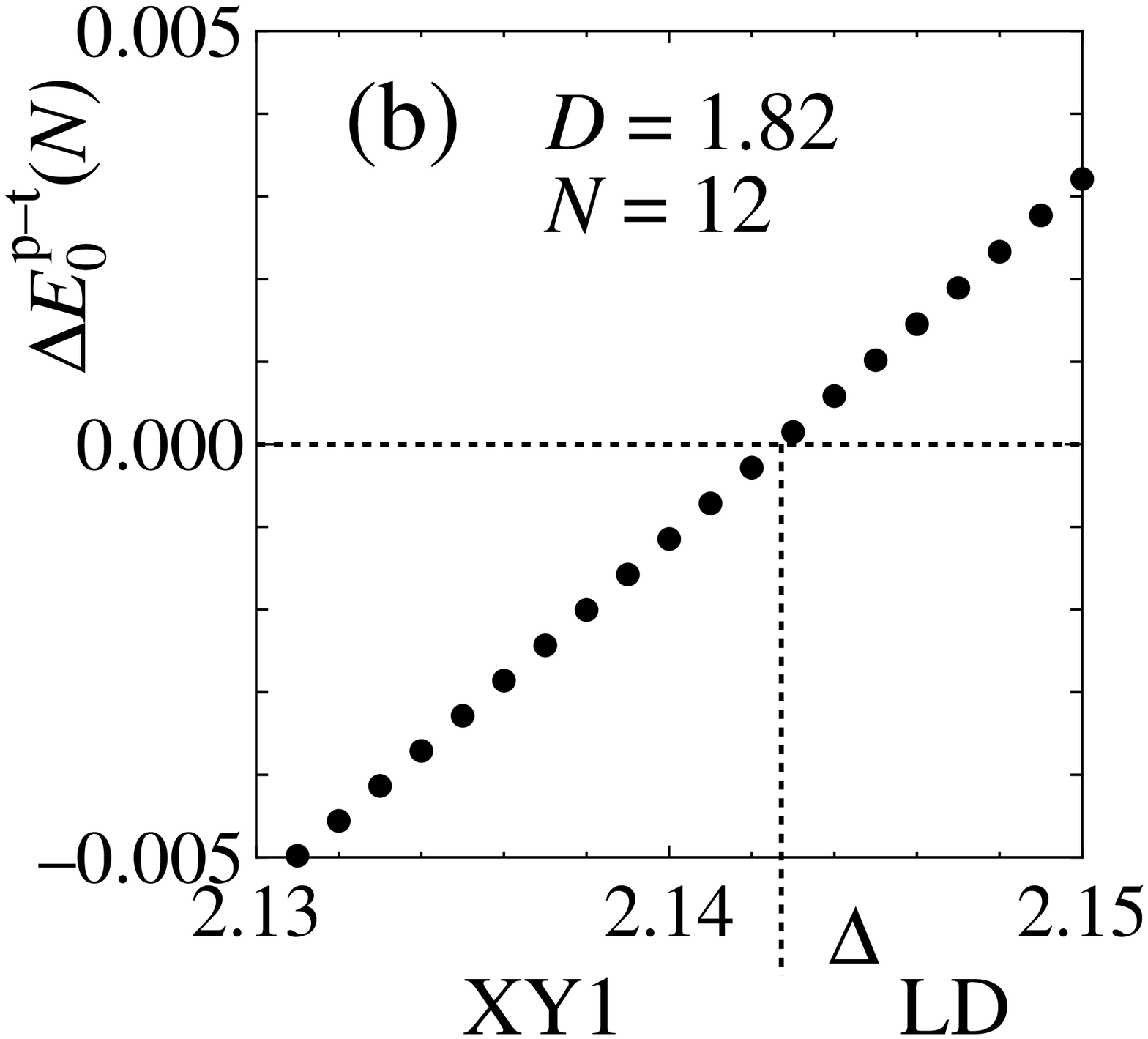}}
   \end{center}
   \caption{(a) Examples of the Haldane-ID and ID-LD transitions
            with \hbox{$D\!=\!1.82$} and \hbox{$N\!=\!12$}, where
            \hbox{$\Delta E_0^{\rm tbc}(N) \equiv E_0(N,0,-1;{\rm tbc})$}
            $-E_0(N,0,+1;{\rm tbc})$.
            From this, we obtain 
            \hbox{$\Delta_{\rm c}^{({\rm H,ID})}(12)\!=\!2.23793$}
            and \hbox{$\Delta_{\rm c}^{({\rm H,ID})}(12)\!=\!2.16639$} by the
            interpolation.  We note that $E_0(N,2;{\rm pbc})$ is higher than
            $E_0(N,0,\pm 1;{\rm tbc})$ in this regime.
            (b) An example of the XY1-LD transition
            with \hbox{$D\!=\!1.82$} and \hbox{$N\!=\!12$}, where
            $\Delta E_0^{\rm p-t}(N) \equiv E_0(N,2;{\rm pbc}) - E_0(N,0,+1;{\rm tbc})$.
            From this, we obtain
            $\Delta_{\rm c}^{(XY1,{\rm ID})}(12) = 2.14265$
            by the interpolation.
            We note that $E_0(N,0,-1;{\rm tbc})$ is higher than
            $E_0(N,0,+1;{\rm tbc})$ in this regime.}
   \label{fig:h-id-n=12a-d=182}
\end{figure}

From the above arguments, for a fixed $D$, the critical value of the Haldane-ID
transition, $\Delta_{\rm c}^{({\rm H,ID})}$, 
and that of the ID-LD transition, $\Delta_{\rm c}^{({\rm ID,LD})}$,
can be evaluated by the \hbox{$N\!\to\!\infty$} extrapolation of
$\Delta_{\rm c}^{({\rm H,ID})}(N)$ and $\Delta_{\rm c}^{({\rm ID,LD})}(N)$,
respectively, determined from the equation\cite{kitazawa,nomura}
\begin{equation}
    E_0(N,0,+1;{\rm tbc}) = E_0(N,0,-1;{\rm tbc}) < E_0(N,2;{\rm pbc}).
    \label{eq:kitazawa}
\end{equation}
Our calculations show that \hbox{eq.~(\ref{eq:kitazawa})} has two
solutions $\Delta_{\rm c}^{({\rm H,ID})}(N)$ and
$\Delta_{\rm c}^{({\rm ID,LD})}(N)$ for \hbox{$N\!=\!8$}, $10$ and $12$ in
the region of \hbox{$1.0\!\lsim \!D\lsim 2.05$}, where
\hbox{$\Delta_{\rm c}^{({\rm H,ID})}(N)\!>\!\Delta_{\rm c}^{({\rm ID,LD})}
(N)$}.  In this region of $D$, we have estimated
$\Delta_{\rm c}^{({\rm H,ID})}(N)$ and $\Delta_{\rm c}^{({\rm ID,LD})}(N)$ by
numerically solving eq.~(\ref{eq:kitazawa}), and then extrapolated these
results for \hbox{$N\!=\!8$}, $10$ and $12$ to \hbox{$N\!\to\!\infty$} by
assuming that their $N$-dependences are quadratic functions of $N^{-2}$.
Examples are shown in Fig.~\ref{fig:h-id-n=12a-d=182}(a) and Table~\ref{table}.
Thus, we have obtained the Haldane-ID and ID-LD boundaries,
which are shown, respectively, by the red and blue
solid lines in Fig.~\ref{fig:phasediagram}.  
\vskip-1.2cm
\begin{table}[h]
\caption{Examples of critical values of $\Delta$ in the case of $D=1.82$ obtained by the LS method.}
\label{table}
\begin{center}
\begin{tabular}{|r|l|l|l|}
   \hline
     N  &$\Delta_{\rm c}^{({\rm H,ID})}(N)$ &$\Delta_{\rm c}^{({\rm ID,LD})}(N)$ &$\Delta_{\rm c}^{(XY1,{\rm LD})}(N)$ \\ \hline    
     6  &2.17687                           &2.14971                            &2.08931  \\ \hline   
     8  &2.22262                           &2.16106                            &2.12246  \\ \hline
    10  &2.23529                           &2.16527                            &2.13607  \\ \hline
    12  &2.23793                           &2.16639                            &2.14265  \\ \hline
    $\infty$ &$2.241 \pm 0.001$             &$2.167 \pm 0.001$                   &$2.156 \pm 0.001$ \\ \hline
\end{tabular}
\end{center}
\end{table}

We have found that the maximum value of $D$ for which eq.~(\ref{eq:kitazawa})
has two solutions exists for each value of $N$ and that, although this maximum
value increases as $N$ increases, it approaches a finite value in the limit of
\hbox{$N\!\to\!\infty$}.  For example, eq.~(\ref{eq:kitazawa}) has two
solutions only when \hbox{$D\lsim 2.05$},
\hbox{$D\lsim 2.14$} and \hbox{$D\lsim 2.17$}, respectively, for
\hbox{$N\!=\!8$}, $10$ and $12$.
We have also found that, when
eq.~(\ref{eq:kitazawa}) has no solution, the relation
\hbox{$E_0(N,M,+1;{\rm tbc})\!<\!E_0(N,M,-1;{\rm tbc})$} always
holds.  These show that, for sufficiently large values of $D$, the ID state
does not appear in the ground-state phase diagram and the crossover between the
Haldane and LD states may take place, which means that the Haldane and LD
states belong to the same phase.  In fact, the extrapolated lines of the phase
boundary line between the Haldane and ID states and that between the ID and LD
states, which are shown, respectively, by the red and blue dotted lines in
Fig.~\ref{fig:phasediagram}, merge at the point
\hbox{$(\Delta, D)\!\sim\!(2.64, 2.19)$}.  

The fact that the Haldane and LD
states belong to the same phase is very reasonable for the following reasons:
(A)~as discussed above, \hbox{$P\!=\!+1$} for both states;
(B)~in the case in which open boundary conditions are imposed, it is clear
that there exists no edge state in the LD state $\bigl($see
Fig.~\ref{fig:vbs-pictures}(b)$\bigr)$, and also Pollmann et
al.\cite{pollmann1,pollmann2} showed very recently that two \hbox{$S\!=\!1/2$}
spins left at each edge in the Haldane state are in the
\hbox{$(S_{\rm tot},S^z_{\rm tot})\!=\!(1,0)$} state, which leads to the
no-edge state, at least if \hbox{$\Delta\!>\!1$} and \hbox{$D\!>\!0$} are
satisfied;
(C) Pollmann et al.\cite{pollmann1}
constructed a one-parameter matrix product state which interpolates
the Haldane and LD states without any quantum phase transition.  Although
the Haldane and LD states are apparently different
from each other from the valence-bond pictures in Figs.~1(a) and (c), they
belong to the same phase.  Here, we give two examples for such a
situation.  One is an antiferromagnetic \hbox{$S\!=\!1$} chain
with the bond alternation and the on-site anisotropy,\cite{tonegawa,hida}
in which the LD state and the dimer state belong to the same phase.
The other is an antiferromagnetic \hbox{$S\!=\!1$} two-leg ladder,\cite{todo}
where the Haldane state and the rung-dimer state belong to the same phase
called the four-site plaquette singlet phase.

In a similar way, for a fixed $D$, the critical values
$\Delta_{\rm c}^{(XY1,{\rm H})}$ and $\Delta_{\rm c}^{(XY1,{\rm LD})}$ of the
$XY$1-Haldane and $XY$1-LD transitions, respectively,
can be obtained from
\begin{equation}
    E_0(N,0,+1;{\rm tbc}) = E_0(N,2;{\rm pbc}) < E_0(N,0,-1;{\rm tbc}).
    \label{XY1-H}
\end{equation}
An example is shown in Fig.~\ref{fig:h-id-n=12a-d=182}(b) and
Table~\ref{table}.  For the critical value $\Delta_{\rm c}^{(XY1,{\rm ID})}$
of the $XY$1-ID transition,
we have to solve 
\begin{equation}
    E_0(N,0,-1;{\rm tbc}) = E_0(N,2;{\rm pbc}) < E_0(N,0,+1;{\rm tbc})
    \label{XY1-H-2}
\end{equation}
instead of eq.~(\ref{XY1-H}).  Thus, we have
obtained the phase boundary lines between the $XY$1 state and one of the
Haldane, ID and LD states, which are depicted, respectively, by the brown,
black and cyanic lines in Fig.~\ref{fig:phasediagram}.  

Finally, the phase transition between the Haldane and N\'eel states is of the
2D Ising type, since the $Z_2$ symmetry is broken in the N\'eel state.
In this case the phenomenological renormalization group
(PRG) method\cite{PRmethod} is useful for determining the phase boundary
line.  We have numerically solved the PRG equation,
\begin{equation}
\begin{split}
    (N-2) &\bigl\{E_1(N-2,0;{\rm pbc}) - E_0(N-2,0;{\rm pbc})\bigr\} \\
          &= N \bigl\{E_1(N,0;{\rm pbc}) - E_0(N,0;{\rm pbc})\bigr\}\,,
    \label{eq:PRequation}
\end{split}
\end{equation}
to obtain the solution $\Delta_{\rm c}^{({\rm H,N})}(N)$ for a given value of
$D$. Then, we have extrapolated, to estimate the critical value
$\Delta_{\rm c}^{({\rm H,N})}$, these results for \hbox{$N\!=\!8$},
$10$ and $12$ to \hbox{$N\!\to\!\infty$} by assuming that their $N$-dependences
are quadratic functions of \hbox{$(N\!-\!1)^{-2}$}, and have determined
the phase boundary line as shown by the green line in
Fig.~\ref{fig:phasediagram}.  
It is expected that $D_{\rm c}^{({\rm H,N})}$ approaches
\hbox{$D_{\rm c}^{({\rm H,N})}\!=\!\Delta$}
in the \hbox{$\Delta\!\to\!\infty$} limit.
This is because in the case of the Ising limit, governed by the Hamiltonian
$\calH_{\rm Ising}= \Delta \sum_j S_j^z S_{j+1}^z+D \sum_j (S_j^z)^2$,
when \hbox{$D\!>\!\Delta$} the
ground state of the present system is the \hbox{$S_{j}^z\!=\!0$} state, and
when \hbox{$D\!<\!\Delta$} it is the N\'eel state with
\hbox{$S_{j}^z\!=\!2(-1)^j$}
and that with \hbox{$S_{j}^z\!=\!2(-1)^{j+1}$} state which are degenerate 
with each other.
In the $\Delta \rightarrow \infty$ limit
the transition line between the N\'eel state and the Haldane state
becomes of the first order,
as is expected from the above Ising limit Hamiltonian $\calH_{\rm Ising}$.
Therefore, on the Haldane-N\'eel transition line,
there should exist a special point at which the transition changes from the 
second order to the first order as $\Delta$ increases.  The details of this
point are left for a future study.

The LS method\cite{kitazawa,nomura} is based on the sine-Gordon theory
or, equivalently, the $c=1$ conformal field theory with the perturbation,
where $c$ is the central charge.
Then, in order to check the applicability of the LS method,
we have calculated the central charge $c$ from the finite-size
correction of the ground-state energy per one spin,\cite{cardy,bloete,affleck}
\begin{equation}
    {E_0(N,0;{\rm pbc}) \over N}
    \simeq \lim_{N \to \infty} {E_0(N,0;{\rm pbc}) \over N} - {\pi v_{\rm s}c \over 6N^2}
\end{equation}
where $v_{\rm s}$ is the spin wave velocity.  
The central charge $c$ is defined only in the gapless region, and it becomes
zero in the gapped region if the above formula is formally applied.  Along the
line $D=1.9$, where the ground-state changes
as $XY1$ $\Rightarrow$ LD $\Rightarrow$ ID $\Rightarrow$ Haldane $\Rightarrow$
N\'eel with the increase of $\Delta$ from zero, we found that the values of
{\lq\lq}apparent $c$" are \hbox{$c\!\simeq\!1$} for
\hbox{$0\!<\!\Delta\!<\!2.7$} and \hbox{$c\!\simeq\!0$} for
\hbox{$\Delta\!>\!2.7$}.  This result shows that our system is effectively
described by the sine-Gordon theory, which ensures the applicability of the LS
method and is consistent with the smallness of gaps in the LD, ID and Haldane
states along this line.  Nomura and Kitazawa\cite{nomura} also found that
\hbox{$c\!\simeq\!1$} at the Haldane-$XY$1 and $XY$1-LD transition points on
the \hbox{$\Delta\!=\!1$} line and at the $XY$1-Haldane transition point on the
\hbox{$D\!=\!0$} line. 

In conclusion, we have determined rather precisely the ground-state phase
diagram on the $\Delta$-$D$ plane of the \hbox{$S\!=\!2$} quantum spin chain
described by the Hamiltonian (\ref{eq:ham}), employing various numerical
methods based on the exact-diagonalization calculation.  The resulting phase
diagram (see Fig.~\ref{fig:phasediagram}) consists of the $XY$1 phase, the
Haldane/LD phase, the ID phase and the N\'eel phase.  It
is noted that in the Haldane/LD phase, the crossover between the
Haldane and ID states may occur.  In our opinion, the most significant result
of the present work is that the ID phase appears in the phase diagram.  The
existence of this phase was predicted by Oshikawa\cite{oshikawa92} about
twenty years ago.  After then, considerable efforts by performing DMRG
calculations\cite{aschauer,schollwoeck1,schollwoeck2} were devoted to find
this phase in the \hbox{$S\!=\!2$} case, but fruitful results were not
obtained.  We now think that the reason for this is clear.
In the DMRG method,
the distinctions between the $XY$1 state and the gapped states are made by
extrapolating the excitation gaps of finite-$N$ systems to
\hbox{$N\!\to\!\infty$}.  Namely, the state is determined to be gapless or
gapped depending on whether the extrapolated gap is zero or finite.  Since, of
course, the values of extrapolated gap have numerical errors, it is very
difficult to discriminate between the gapless state and the state with a very
small gap.  This situation is clearly seen in Fig.~7 of the paper of Aschauer
and Schollw\"ock.\cite{aschauer} 
In other words, the apparent boundaries between the $XY$1
phase and the gapped phases determined by the DMRG method shift into the gapped
region, because the system with \hbox{$\xi\!>\!N$} behaves to be gapless,
where $\xi$ is the correlation length.
In fact, in the case of \hbox{$\Delta\!=\!1$}, the transition point between
the $XY$1 phase and the LD phase is estimated as
\hbox{$D_{\rm c2}\!\sim\!3.0$} by the DMRG method,\cite{schollwoeck2}
which is considerably larger
than $D_{\rm c2}=2.39$,\cite{nomura} obtained by the LS
method.  As can be seen from Fig.~\ref{fig:phasediagram}, the region of the ID
phase is fairly narrow, and thus the whole region of the ID phase was
completely masked by the {\lq\lq}apparent $XY$1 region".  Although we have
tried to detect the edge state with \hbox{$S^z\!=\!1/2$} $\bigl($see
Fig.~\ref{fig:vbs-pictures}(b)$\bigr)$ around each end of the open chain in
the ID state by the DMRG method, this has also led to fruitless
results.  The reason why the LS method \cite{nomura,kitazawa} is very
useful is undoubtedly the fact that it can sharply distinguish three phases,
the $XY$1, ID and Haldane/LD phases, by the crossing of the excitations under
periodic and twisted boundary conditions.

We would like to express our appreciation to Professor Masaki Oshikawa for his
invaluable discussions and comments as well as for his interest in this
study.  We are deeply grateful to Dr. Frank Pollmann for his stimulating
discussions.  We also thank the Supercomputer Center, Institute for Solid
State Physics, University of Tokyo, the Computer Room, Yukawa Institute for
Theoretical Physics, Kyoto University and the Cyberscience Center,
Tohoku University for computational facilities.  The present work has
been supported in part by a Grant-in-Aid for Scientific Research (B)
(No.~20340096), and a Grant-in-Aid for Scientific Research on Priority Areas
^^ ^^ Novel states of matter induced by frustration'' from the Ministry of
Education, Culture, Sports, Science and Technology.

\end{document}